\documentclass [12pt,twoside]{article}           
\usepackage[left,modulo]{lineno}
\usepackage{setspace}

\countdef\arxiv=201

\ifnum\arxiv=0 \input cpreb.tex \fi

\ifnum\arxiv=1

\usepackage{epsfig,times,lscape}
\usepackage[usenames]{color}

    \ifnum\count102=1

    \fi

    \ifnum\count102=2
\fi

\newcommand{\nc}{\newcommand}

     \ifnum\count101=1
\nc{\qI}[1]{\section{{#1}}}
\nc{\qA}[1]{\subsection{{#1}}}
\nc{\qun}[1]{\subsubsection{{#1}}}
\nc{\qa}[1]{\paragraph{{#1}}}

\def\qpar{\vskip 2mm plus 0.2mm minus 0.2mm}
\def\qL{\hfill \break}
     \fi 

      \ifnum\count101=2
 \nc{\qI}[1]{\parindent=0mm \vskip 8mm 
{\centerline{\LARGE \color{red}#1}}\vskip 3mm}
\nc{\qA}[1]{\vskip 2.5mm \noindent 
{{\bf\large\color{blue}  #1}} \vskip 1mm \parindent=0mm}
 \nc{\qun}[1]{\vskip 1mm \noindent {\sl #1 }\quad }

\def\qL{\hfill \break}
\def\qpar{\vskip 2mm plus 0.2mm minus 0.2mm}

      \fi

\def\qth{\vrule height 12pt depth 0pt width 0pt}
\def\qtb{\vrule height 0pt depth 5pt width 0pt}

\nc{\qfoot}[1]{\footnote{{#1}}}

      \ifnum\count101=1
\def\qbu{\hfill \par \hskip 6mm $ \bullet $ \hskip 2mm}
\def\qee#1{\hfill \par \hskip 6mm (#1) \hskip 2 mm}
      \fi
      \ifnum\count101=2
\def\qbu{\hfill \par \hskip 4mm $ \bullet $ \hskip 2mm}
\def\qee#1{\hfill \par \hskip 4mm (#1) \hskip 2 mm}
      \fi

\def\qparr{ \vskip 1.0mm plus 0.2mm minus 0.2mm \hangindent=10mm
\hangafter=1}

     \ifnum\count101=1 
 
     \fi
     \ifnum\count101=2

  \def\qcitb#1{\noindent \hbox to 102mm{\hfill \small #1} \vskip 1mm}
      \fi

 \def\qpages#1{\count102=0{\loop\advance\count102 by 1
 \null \vfill\eject \ifnum\count102<#1 \repeat}}

\def\qn#1{\eqno \hbox{(#1)}}

\def\qth{\vrule height 12pt depth 0pt width 0pt}
\def\qtb{\vrule height 0pt depth 5pt width 0pt}

\def\qv{\vskip 0.1mm plus 0.05mm minus 0.05mm}
\def\qhu{\hskip 0.6mm}
\def\qhv{\hskip 3mm}

\def\qhw{\hskip 1.5mm}
\def\qleg#1#2#3{\noindent {\bf \small #1\qhw}{\small #2\qhw}{\it \small #3}\qv }
\newcommand{\promille}{%
  \relax\ifmmode\promillezeichen
        \else\leavevmode\(\mathsurround=0pt\promillezeichen\)\fi}
\newcommand{\promillezeichen}{%
  \kern-.05em%
  \raise.5ex\hbox{\the\scriptfont0 0}%
  \kern-.15em/\kern-.15em%
  \lower.25ex\hbox{\the\scriptfont0 00}}

\fi

\begin{document}
\thispagestyle{empty}



\markboth{{\sl \hfill  \hfill \protect\phantom{3}}}
        {{\protect\phantom{3}\sl \hfill  \hfill}}

\color{yellow} 
\hrule height 10mm depth 10mm width 170mm 
\color{black}

 \vskip -17mm   

\centerline{\bf \Large Coupling between death spikes
and birth troughs}
\vskip 5mm
\centerline{\bf \Large Part 2: Comparative analysis of salient 
features}
\vskip 10mm

\centerline{\large 
Peter Richmond$ ^1 $ and Bertrand M. Roehner$ ^2 $
}

\vskip 10mm
\large

%
{\bf Abstract}\qL
In part 1 we identified a new coupling between death spikes and birth dips that
occurs following catastrophic events such as influenza pandemics and
earthquakes.
Here we seek to characterise some of the key features. We introduce a transfer
function defined as the amplitude of the birth trough (the output)
divided by the
amplitude of the death spike (the input). This has two features: it is always
greater than one so is an attenuation factor and as a function of the
amplitude of
the death spike, it may be characterized by a power law with exponent close to
unity. Since many countries do not publish monthly data, merely annual
data, we
attempt to extend the analysis to cover such data and how to
identify the death-birth
coupling. Finally we compare the response to unexpected death
spikes and regular seasonal death peaks, such as winter death peaks
which occur in many countries.

\vskip 10mm
\centerline{\it \small Version of 14 October 2017.}
\vskip 5mm

{\small Key-words: death rate, birth rate, shock, 
transfer function, seasonal pattern.}

\vskip 5mm

{\normalsize
1: School of Physics, Trinity College Dublin, Ireland.
Email: peter\_richmond@ymail.com \qL
2: Institute for Theoretical and High Energy Physics (LPTHE),
University Pierre and Marie Curie, Paris, France. 
Email: roehner@lpthe.jussieu.fr
}

\vfill\eject

\qI{Introduction}

The case-studies described 
in the previous paper (Richmond et al. 2017, thereafter
referred to as ``Paper 1'') specified some of the conditions
which must be fulfilled for this effect to exist. The fact
that it takes place for the H1N1 crisis in Hong Kong
but not for the attack of 9/11 in New York led to the idea
that it is not really the number of deaths which is the
main determinant, but rather the total number of persons
who experience an adverse shock in their living
conditions. 
\qpar

In the present paper we have three objectives.
\qee{1} In Paper 1 the coupling effect was represented
(in Fig. 2a) as an input-output system. It is therefore
natural to measure the transfer function of this system. 
In particular we wish to see if the system is linear
or nonlinear.
\qee{2} Secondly, we wish to extend the analysis of the
coupling effect to cases for which only annual data are
available. This would represent a significant extension for
monthly data are unavailable in many developing
countries, either because they are collected but not sent
to the central government or because the central
government gets them but does not publish them.
\qee{3} Apart from the exceptional death spikes due to special
events, monthly mortality data display also seasonal
peaks. The amplitudes of such peaks are country-dependent and
in some countries they reach levels which are as high or even higher
than the exceptional death spikes. It is therefore natural
to compare their respective effect on birth numbers.

\qI{Attenuation factor as a function of death spike amplitude}

In Paper 1, it was suggested that the main determinant
is the number of persons
who experience an adverse shock in their living
conditions.
 Unfortunately, in many cases this number
is not well defined. For instance 
the measure of the incidence of a disease 
is highly dependent upon the criterion that is used:
\qbu The number of persons hospitalized gives a 
low measure of incidence.
\qbu A broader measure of incidence is through the number of 
working days lost to sickness in the labor market.
\qL
However, statistical data corresponding to these criteria are
rather sparse and not comparable across a set of countries.
\qpar

In the case of earthquakes we suggested
that the shock on survivors could
be measured through the number of ``damaged houses'' but this 
latter notion is itself a matter of appraisal.
\qpar

For all these reasons the number of deaths remains the 
most convenient parameter for it has a clear significance
and is widely available in vital statistical records.

\qA{Method}
 
As for all time series which show a seasonal pattern
we need to resolve how to handle it.
The methods that 
we will use successively 
rely on two different conceptions of the phenomenon under
consideration. 

\qun{All inclusive conception (1)}\qL
In the first conception we consider the death spike as being of the
same nature as the seasonal fluctuations. In other words it is
seen as a seasonal fluctuations which just happens to be
somewhat higher than the others. In this conception it would
not make sense to separate the two effects. This means that we
measure the amplitude of the death spike (and similarly for
the birth trough) just ``as it is''. The beginning of the spike
will be defined as the month where the number of deaths starts
to increase after having been decreasing or flat. Similarly, the
end of the  spike will be the month where the deaths start to
level off or to increase. Naturally, even if there is a 
small local dip
in the upward phase or a local surge in the downward phase
we do not wish them to be taken into account. That is
why we perform a 3-point centered moving average before 
implementing the previous procedure. 
\qpar

%
\begin{figure}[htb]
\centerline{\psfig{width=14cm,figure=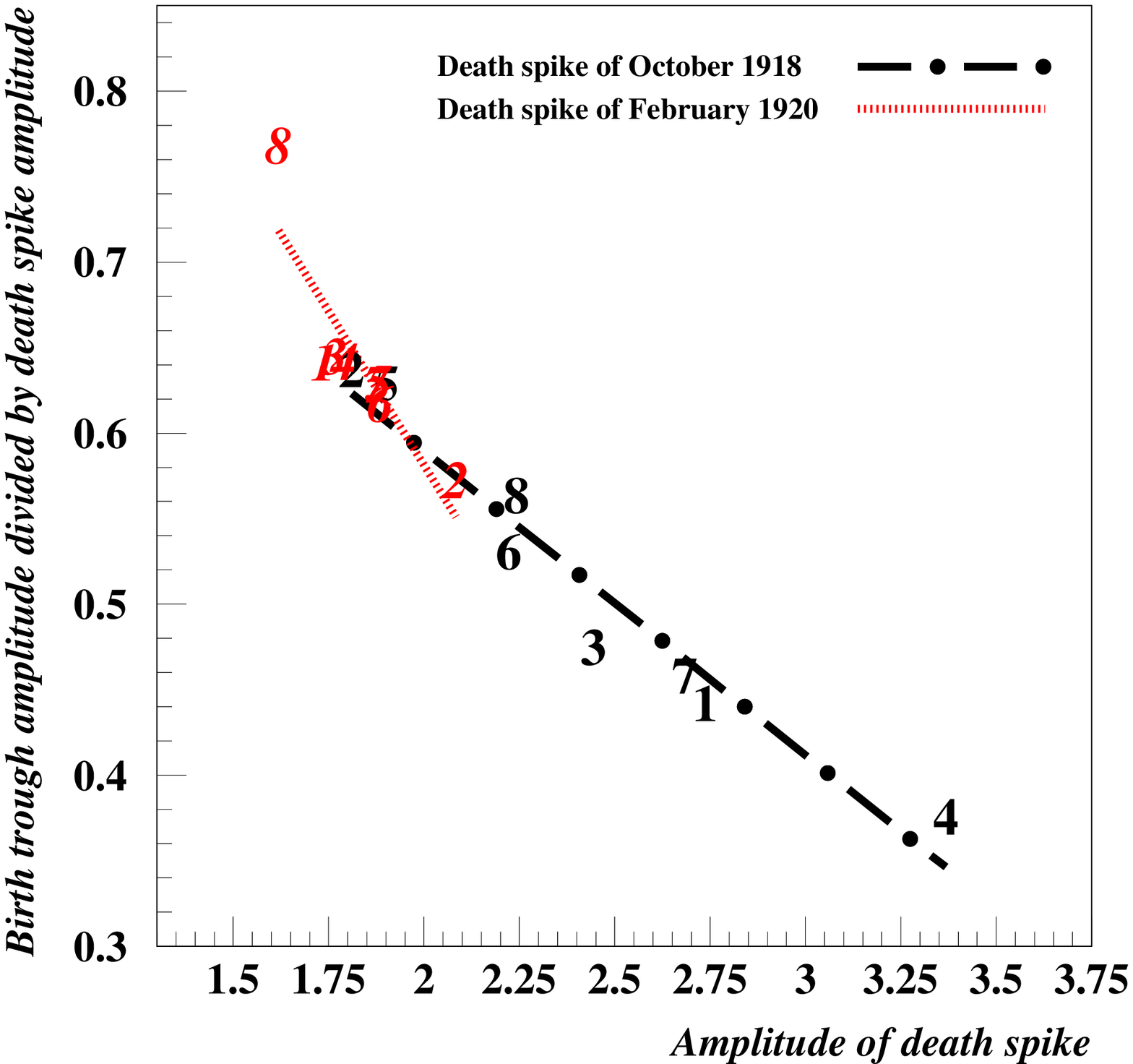}}
\qleg{Fig.\qhu 1\qhv Relationship between the 
the influenza death spikes of 1918 and 1920 in the United
States and subsequent birth troughs.}
{The graph describes the function: $ R=A_b/A_d=f(A_d) $
where: $ A_d\ =\ $amplitude of death spike, 
$ A_b\ =\ $amplitude of birth trough.
As the death spike of 1920 was markedly smaller
than the one of 1918 it permits an exploration of the
small $ A_d $ section; this exploration
suggests that the function $ R=f(A_d) $
is probably nonlinear. The meaning of the numbers is as
follows: 1=Massachusetts, 2=Michigan, 3=New York, 4=Pennsylvania,
5=Indiana, 6=Ohio, 7=Cities of the Registration Area, 8=Rural
parts of the Registration Area. The regressions read as follows
(the confidence
intervals are for a confidence level of 0.95):\qL
1918: $ R=aA_d+b,\ a=-0.18\pm 0.04,\ b=0.94\pm 0.02 $
(correlation=$ -0.97 $).\qL
1920: $ R=aA_d+b,\ a=-0.36\pm 0.15,\ b=1.3\pm 0.02 $
(correlation=$ -0.88 $).\qL}
{Sources: Bureau of the Census: Mortality Statistics 1917--1921;
Bureau of the Census: Birth Statistics 1917--1921.}
\end{figure}

%
\begin{figure}[htb]
\centerline{\psfig{width=17cm,figure=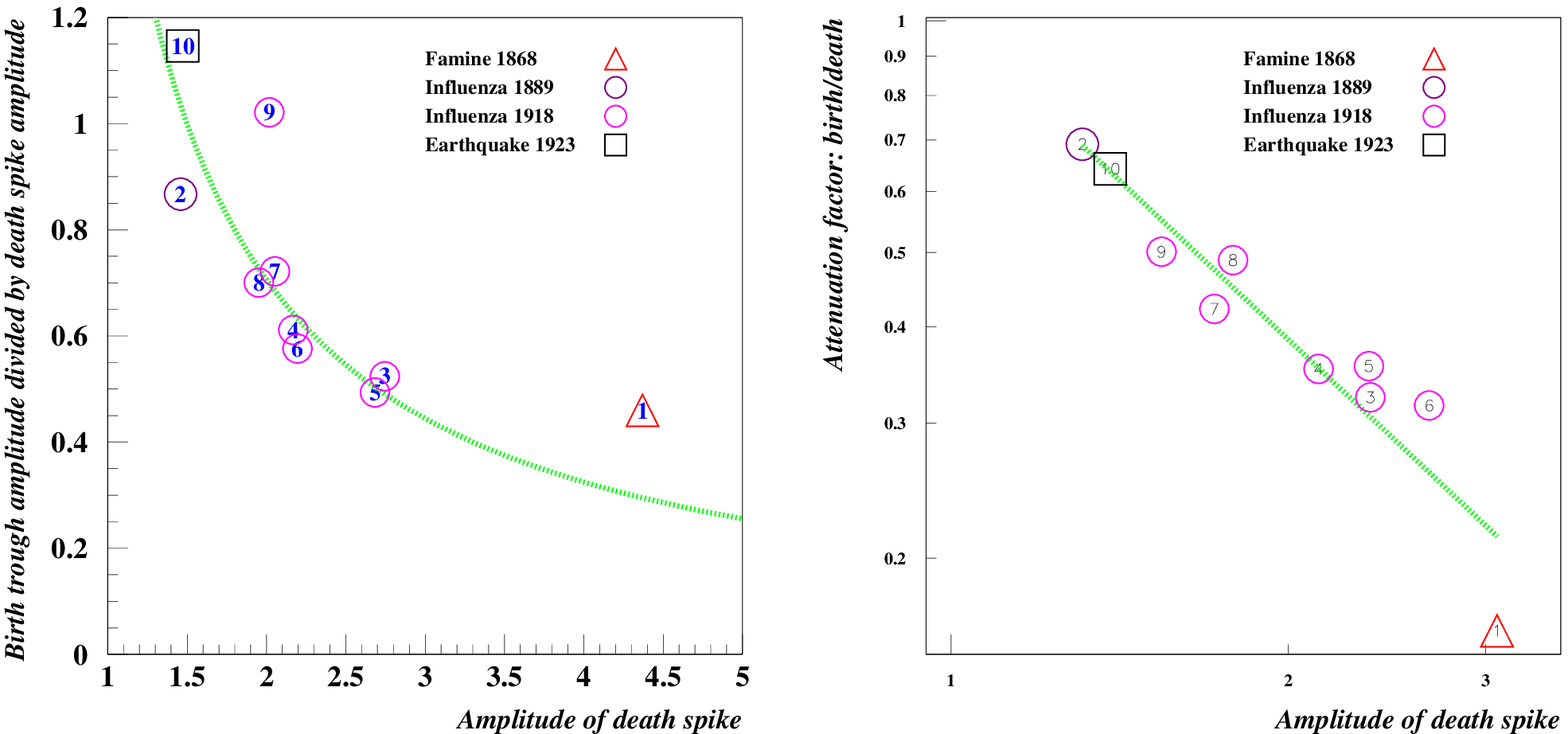}}
\qleg{Fig.\qhu 2a,b\qhv Relationship between 
the amplitude of death spikes and the attenuation factor
birth/death.}
{Both graphs represent the ratio $ R=A_b/A_d $
where: $ A_d\ =\ $amplitude of death spike, 
$ A_b\ =\ $amplitude of birth trough. The graphs differs
by the method used in the calculation. Both results are
compatible with a power law relationship. In Fig. 2b
this relationship is shown in log-log scales. 
The term 
``attenuation factor'' expresses the fact that the ratio $ R $
is smaller than one; the point higher than 1 in Fig 2a
may be a statistical fluctuation due to the fact that
the smaller the amplitudes the more fluctuating their
estimates. The two graphs rely on different conceptions
of the fluctuations
that are explained in the text. Overall, they lead to
similar results, namely: $ R\sim 1/A_d^{\alpha} $
where $ \alpha $ is of the order of 1.
The linear regression estimates for the logarithms,
read $ \log(R)=\alpha \log(A_d)+b $ with the following estimates
for the parameters (confidence level is 0.95).
Fig 2a: $ \alpha=-0.81\pm 0.4,\ b=0.25\pm 0.11 $ (correlation
is $ -0.83 $ ; Fig. 2b:
$ \alpha=-1.37\pm 0.4,\ b=-0.001\pm 0.1 $ (correlation is $ -0.94 $). 
The two estimates of $ \alpha $ are compatible with $ \alpha\simeq 1 $. 
The plotted numbers correspond to the following cases:
1:Finland 1868 (famine); 2:France 1889 (influenza); 3-9:influenza
epidemic in several countries which did not take part in
World War I:
Sweden (3), Switzerland (4), Spain (5), Denmark (6), Finland (7),
Chile (8), Japan (9); 10:Tokyo 1923 (earthquake and fire).
}
{Sources: Bunle (1954), Finland (1902). Statistique de la France,
Nouvelle s\'erie (various years).}
\end{figure}

\qun{Seasonal fluctuations seen as noise (2)}
\qL
In the second conception in which one considers that the death spike is
of a different nature than the seasonal fluctuations, the challenge
is to remove the seasonal variations in the ``best'' possible way.
In principle, the way to do that seems fairly evident and consists
in dividing the monthly deaths of year $ y_0 $ by the seasonal profile
that we denote by $ P_s $ (it is a set of 12 numbers).
But how should the seasonal profile be defined? The answer depends
upon the characteristics of the seasonal pattern.
The simplest way is to take the monthly death profile of the year $ y_{-1} $
preceding $ y_0 $, in other words: $ P_s=D(y_{-1}) $.
The main advantage of such a choice is the
fact that in case there is  a drift of the seasonal profile
in the course of time, the year closest to $ y_0 $ will be the 
most appropriate.  
\qpar

A possible drawback of taking $ D(y_{-1}) $ is the fact that, as
a single year, it may differ from the average seasonal pattern.
Instead of taking only one year it is tempting to think
that an average over several years would better approximate $ P_s $.
Is that true?\qL
If the inter-annual statistical fluctuations of seasonal variations
are small, then the average of $ n $ years
will indeed converge toward a reasonable seasonal pattern.
However, one should observe that in such a case $ D(y_{-1}) $ 
differs little from the average and is also a good
choice therefore. \qL
On the contrary, if from year to year there are large random
changes in the monthly pattern, then an average of several years
will be almost flat and the more years one takes the flatter
it will become%
\qfoot{In order to discuss this point theoretically one would have 
to know the statistical frequency functions of the deaths (or births)
in each months and also the interdependence of deaths in neighboring
months. The statement that the average of several years
tends to become level relies on tests performed for the
Japanese death data which is one of the most regular.}%
. 
Such an average will be useless therefore
and in such a case $ D(y_{-1}) $ will probably be the best
choice as being close to $ y_0 $. 
\qpar

In summary  we retain two procedures:
\qee{1} Scaling the spikes and troughs just ``as they are''.
\qee{2} Scaling them after dividing them by $ D(y_{-1}) $ and
 $ B(y_{-1}) $ respectively. 
\qpar

In what follows we will try successively the two procedures.

\qA{Selection of the data samples}

Here, again, there are two different strategies.
\qee{i} One can select an homogeneous sample of cases, 
such as the 1918 influenza epidemic in the US. 
This has the advantage of good comparability but the
drawback of a fairly narrow interval for the
amplitudes $ A_d $ of the death spikes.
\qee{ii} One can consider a broad set of cases
which includes famines, diseases, earthquakes,
terrorist attacks. This has the advantage of a wider
range for the death spike amplitude, but the drawback
of increasing the noise by mixing different
kinds of cases.
\qpar

In what follows we will try both strategies. 
In strategy (i) the range of $ A_d $ will be $ (1.5,3.3) $
whereas in strategy (ii) it will be extended to $ (1.5,4.5) $.

\qA{The influenza pandemic in the United States}

In the United States the development of the statistical network
was slower than in smaller and more centralized countries
like France or Sweden. Death statistics were recorded in the
so-called Death Registration Area whereas birth data
were recorded in the Birth Registration Area.
In 1917 there were only 19 states in the Death Registration
Area. However, some of them did not belong to the
Birth Registration Area. That was for instance the case
of California; as we need both death and
birth data, California could not be used in our investigation.
In addition we omitted a number of small states such as Delaware,
New Hampshire or Rhode Island because for small
states the monthly death
numbers would be too low and therefore would show
large fluctuations. That is why our sample comprises only 8 cases.
\qpar

In the US, the war has had little influence on married
couples because husbands belonged to class
IV of the ``Selective Service System'' which means that 
they were drafted only after the resources of the
classes I, II, III had been exhausted. In short, 
one can admit that only a small percentage of husbands
were drafted. Naturally, as can be expected, this rule
incited many young people to get married to avoid
the draft. From 1916 to 1921, according to Bunle (1954, p. 257)
the numbers of marriages in Massachusetts
were as follows (in thousands): 
$$ \matrix{
\qtb
1916 & 1917 & 1918 & 1919 & 1920 & 1921 \cr
\noalign{\hrule}
\qth 
34.3 & 37.9 & 29.1 & 34.3 & 38.0 & 33.5 \cr
}
$$ 

The data confirm that there was a marriage surge
in 1917 and this effect is further confirmed by the monthly data.
The United States declared war on 
Germany on 6 April 1917; in the 3 months Jan-Mar the 
marriages were almost the same in 1917 as in 1916 but
in the quarter Apr-Jun they were 30\% higher.
\qpar

Fig. 1 shows that the level of noise is sufficiently small
for a well-defined relationship to exist  between $ A_d $ and
the attenuation ratio $ R=A_b/A_d $. The points 7 and 8
refer to urban and rural parts respectively 
and the fact that they are in line with the other points shows that
the urban/rural factor does hardly affect the $ R=f(A_d) $ relationship.
\qpar
Not surprisingly, the level of noise is somewhat larger for
the smaller death spikes of 1920 than for the spikes of 1918.

\qA{Broad sample of various death spikes}

The small level of noise experienced in the previous data set
encourages us to try a broader one. Fig. 2a shows
a higher dispersion of the data points especially for the
smallest death spikes but the level of noise remains
acceptable.
\qpar

Fig. 1 and Fig. 2a were made with methodology (1), whereas
Fig. 2b was made with methodology (2); it can be seen that it is
the case of Finland which is most affected but overall the 
two methods lead to similar results. 

\qA{Nonlinearity of $ R=f(A_d) $}

Fig. 2a and 2b show clearly that, as already suspected
in Fig. 1, the relationship  $ R=f(A_d) $ is not linear.
Big death spikes have a lower attenuation factor
than small ones but it seems to converge toward
a limit. 
\qpar

A simple interpretation can be given.
In paper 1 we have seen that in the earthquake of 2011 in
Japan, some 400,000 houses were damaged. If we take this
number as representing the persons directly affected 
the ratio to the number of deaths will be $ M=400,000/18,000=22 $.
Now, in the case of Finland in 1868, the number of excess deaths
was $ 80,000 $. By applying the same multiplier $ M=22 $
we get a number of $ 80,000 \times 22 = 1.8 $ million.
However, this number is equal to the whole population of
Finland in 1868 which is unrealistic because
wealthy persons were certainly not affected by the famine.
In short, the large multipliers which are possible for small
death spikes are not possible for large death spikes 
simply because of the limit imposed by the number of
persons exposed to the risk.

\qI{Can one use annual instead of monthly birth-death data?}

The investigation of the death-birth coupling
requires high frequency (monthly or weekly)
data; however when such data are unavailable
the coupling can, under appropriate conditions, be identified
through its specific signature at annual level.
This is the point which will be discussed in this section.

\qA{Motivation}

Although the Statistics Division of the United Nations publishes
monthly birth and death data for many countries, there are
quite a few important countries (e.g. China, India, Indonesia,
Thailand) for which such statistics are not available.
Even for countries included in the list the data are missing
for some years. 
This raises the question of whether the pattern visible at
the monthly level also results in a recognizable signature
at annual level. If so, that would allow us to extend
our analysis  to a number of cases for which no
monthly data are available; examples are 
the Tangshan earthquake in northeastern China on 28 July 1976,
4am (about 250,000 deaths),
the Indian Ocean tsunami of 26 December 2004, 8am (250,000 deaths)
the Great Sichuan
earthquake in west China on 12 May 2008, 2:30pm (90,000 deaths).
\qpar

The fact that for annual data there are no seasonal fluctuations
should be a favorable factor but 
we first need to compare monthly and annual 
fluctuations of birth numbers.

\qA{Monthly versus annual fluctuations of birth numbers}

\begin{table}[htb]

\small

\centerline{\bf Table 1 \quad Monthly and annual fluctuations 
of birth numbers}

\vskip 5mm
\hrule
\vskip 0.7mm
\hrule
\vskip 2mm

$$ \matrix{
\hbox{} \hfill & \hbox{1} & \hbox{2}& \hbox{3}\cr
\hbox{} \hfill & \hbox{Coeff. of} & \hbox{Average of}& \hbox{Stand. dev.}\cr
\qtb
\hbox{} \hfill & \hbox{variation} & \hbox{abs. changes}& \hbox{of
  logs}\cr
\noalign{\hrule}
\qth
\hbox{Monthly births} \hfill & 6.8\%& 6.5\%& 6.8\% \cr
\hbox{Annual births} \hfill & 4.2\%& 3.8\% & 4.2\% \cr
\hbox{} \hfill & & & \cr
\qtb
\hbox{Ratio monthly/annual} \hfill & 1.60 & 1.70 & 1.61\cr
\noalign{\hrule}
} $$
\vskip 1.5mm
Notes: The coefficient of variation is the ratio: standard
deviation/mean. The second column gives the average of the absolute
values of successive relative changes. The third column gives the
standard deviation of the logarithms of birth numbers.
The fact that the ratio monthly/annual is equal to 1.60 instead
of $ \sqrt{12}\simeq 3.5 $ is due to the autocorrelation of the
monthly birth data (see text).
The data are for Sweden; the monthly data cover the 10 years
Jan 1911--Dec 1920 (divided into 5 series of 2 years) while the
annual data cover the 100 years 1821-1920 (divided into 10 series
of 10 years).\qL
{\it Sources: Bunle (1954), Flora et al. (1987)}
\vskip 2mm
\hrule
\vskip 0.7mm
\hrule
\end{table}

Estimates of the fluctuations are given in Table 1. 
The data are for Sweden but are certainly similar for
other countries.
Instead of the rates we considered the numbers
of births because in the early 19th century the total population
was probably known with less accuracy than the birth numbers.
Moreover,
in order to avoid the bias due to the downward trend (related
to the demographic transition) the global series were split into
10 annual series and 5 monthly series.
\qpar
The three estimates considered in Table 1 are related but are
not equivalent. Although not a standard one, estimate 2 is the most
transparent for the present purpose.

As each annual value is the sum of 12 monthly numbers one would expect
a coefficient of variation which is smaller by a factor 
$ \sqrt{12}=3.46 $. Why then does it turn out to be only 1.60 times
smaller? It is related to the fact that successive changes
are not independent. The standard deviation $ \sigma (n) $ of
the average of $ n $ random
variables of standard deviation $ \sigma $ and
whose pair-wise correlation is on average equal to
$ r $ is given by the formula%
\qfoot{The proof is straightforward and recalled in 
Roehner (2007, p. 45)}%
:
$$ \sigma (n)={ \sigma \over \sqrt{n} }g,\quad g=\sqrt{1+(n-1)r} $$

For independent variables one gets the standard result: 
$ \sigma/\sigma (n) =\sqrt{n} $. 
Here, with $ n=12 $ and 
$ g=\sqrt{n}/[\sigma/\sigma (n)]\simeq 3.46/1.6\simeq 2.1 $ one gets $
r\simeq 0.33 $.
Is this prediction consistent with the values given by the autocorrelation
function $ \rho_j $ where $ j $ is the time lag expressed in months?\qL
\qbu A rough test is to observe that $ \rho_6 $ is of the order of
$ 0.3 $.
\qbu For a more accurate test one needs to compute the average
of the correlations of all pairs of months.
Naturally, the number of pairs depends upon the time-lag.
For a time lag of 1 month there are 11 pairs ($ 1-2,2-3,\ldots ,11-12 $), 
whereas for a time lag of 10 there are only 2 pairs ($ 1-11,2-12 $).
Altogether one gets:
$ r_p=(1/66)\Sigma_1^{11}(12-j)\rho_j $. Plugging in the
values of the autocorrelations, one gets: $ r_p=0.36 $
which is consistent with the value of $ r $ predicted above.

\qA{Conditions under which one expects a recognizable signature}

At monthly level one observes the following succession of events:\qL
The death spike is followed 9 months later by a birth trough which is
itself followed 3 or 4 months later by a birth rebound
(the rebound effect is documented in Paper 1).
It is the predicted succession of these events which helps us
to identify them. Under what circumstances can one expect
a similar pattern for annual data?
\qbu
Consider a death spike which occurs in November of year $ y_0 $.
The minimum of the birth trough would be expected in July
of year $ y_1=y_0+1 $. 
The rebound would start about 4 months later, that is to
say in November of $ y_1 $; however, most of it will occur 
in the following year, namely $ y_2=y_0+2 $. 
This is the ideal case because it results
in a well staged succession of yearly death and birth levels:
 $$ d(y_{-1})<d(y_0)=\hbox{\small spike} >d(y_1) 
\quad b(y_0)>b(y_1)=\hbox{\small trough} <b(y_2)=
\hbox{\small rebound} \qn{\hbox{A}} $$

Such cases will be referred to as ``class A'' cases. 
If the death spike occurs earlier in $ y_0 $ the situation will
be less favorable. 
\qbu If the death spike occurs between
January and April the birth
trough will take place (partly or totally) in $ y_0 $ and the rebound
will take place in  $ y_1 $. This leads to the following signature
which will be referred to as ``class B''.
 $$ b(y_{-1})>b(y_0)=\hbox{\small trough}<b(y_1)=
\hbox{\small rebound}>b(y_2)\qn{\hbox{B}} $$
\qbu If the death spike occurs between May and October
the birth trough will be between February and July of $ y_1 $.
In this case whether $ b(y_1) $ will be lower or
higher than $ b(y_0) $  will depend upon how fast the rebound starts
and how strong it is. This mixed and fairly unclear situation 
will be referred to as ``class AB''.
\qpar

In summary, one can remember the following rules.
(i) If the death spike occurs
in January or February one expects: $ b(y_0)<b(y_1) $.
(ii) When the death spike occurs in March or
April one is in a mixed situation for which one does not expect
any clear relationship for birth numbers. 
(iii) If the death spike
occurs between May and December 
one expects: $ b(y_0)>b(y_1)<b(y_2) $

\qA{How to use annual data?}

%
\begin{figure}[htb]
\centerline{\psfig{width=16cm,figure=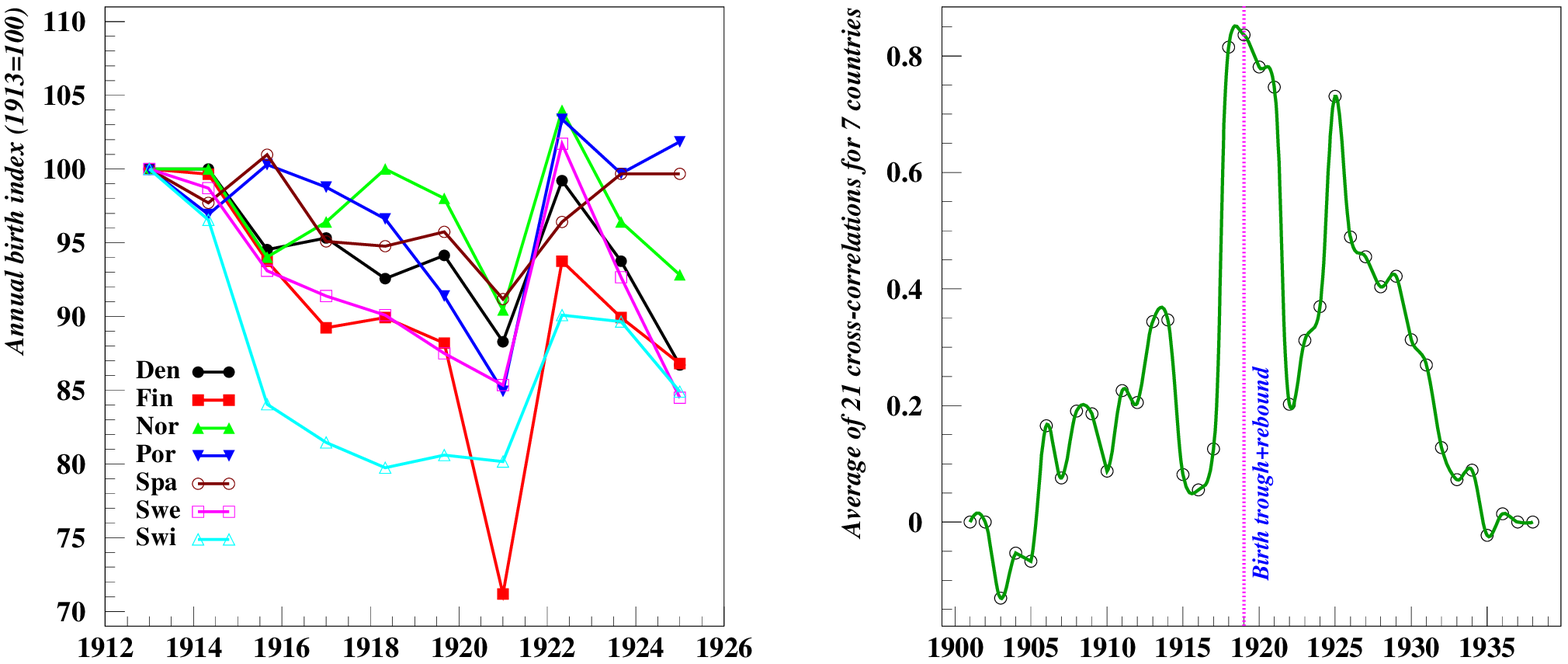}}
\qleg{Fig.\qhu 3a,b\qhv Identification of systemic birth
events within a set of countries.}
{The graph on the right-hand side shows that the trough
and rebound of 1919-1920 is the only major systemic birth event 
in the time interval 1900-1938. 
This is confirmed by the average cross-correlation shown
in Fig. 3b.
The graph 3b was made with a
moving window with a width of 5 months.}
{Source: Mitchell (1978)}
\end{figure}

We now come to the most important part of this section.
How can we apply what we have learned about annual birth data
in order to make them into a useful tool?
From the discussion above we know that we should select events
which occurred toward the end of year. As the 1918 influenza pandemic
in the northern hemisphere occurred in October-November it would be a 
good candidate. We know that the birth trough will be located
sometime around July 1919 but with annual data it will be 
spread over the whole year and therefore become ``diluted'' about
6 times (if the monthly trough lasts two months). On the
other hand the background noise will be reduced only
by as factor 1.6. Thus the identification will be 3.7 times
more difficult.
This leads us to work
in a statistical perspective that is to say by exploring
a whole set of countries as done in Fig. 3b.
\qpar

The set of countries consists of 7 European countries,
none of which took part in the First World War.
Fig 3a shows that their birth fluctuations are fairly 
disconnected
except for two changes which are common to most of them, namely
the dip of 1919 and the rebound of 1920.
This widespread accident appears as a correlation peak in
Fig. 3b. This graph was made by computing the
cross-correlations of the 21 pairs of countries over a moving window
and then summing them up. The correlations add together
destructively 
except in the interval around 1919-1920
and in a narrow interval around 1925. 
\qpar
In short, this method permits to identify collective
motions of birth rates. 
\qpar
Can we repeat for the Indian Ocean
Tsunami of 2004 the identification operation done in 
Fig. 3b? The answer is ``no''. The reason is simple.
The 3 countries with the highest death rates were
Sri Lanka (35,000 deaths, i.e. 1.7 per 1,000), 
Indonesia (131,000 deaths, i.e. 0.65 per 1,000) and
Thailand (5,400 deaths, i.e. 0.09 per 1,000). So, there
were only two countries with death rates over 0.1 per 1,000.
Moreover,
for one of them, namely Indonesia, there are no annual
birth and death data
reported in the Demographic Yearbook of the United
Nations.

\qI{Exceptional versus seasonal death upsurges}

Since exceptional death upsurges as those considered so far 
give rise to birth troughs, should one not expect similar
responses for the recurrent upsurges of seasonal mortality?
This is a question which comes about naturally and must therefore
be addressed. However, we will see that it is not a well defined
question in the sense that its answer is country dependent.
This is due to the fact that, apart from the Bertillon effect,
the fluctuations of birth numbers are also influenced by 
other factors, for instance climatic features as well as
cultural and religious rules.
This can be seen fairly clearly in the case of
Japan by the following observations.
\qbu In 1914-1917 the coefficient of variation (CV=standard deviation
divided by average) is equal to 9.4\% for the deaths and 29\% for
the births. As we have seen previously that the Bertillon effect
is an attenuation (not an amplification) the fact that CV(birth) is
three times CV(death) shows that there are exogenous factors at work.
\qbu The previous argument is comforted by the following
observation. Between 1906 and 2013 CV(death) remained fairly
constant around 10\% whereas CV(birth) fell from 30\% to 3.2\%.
This suggests a decline of the exogenous factors in the course
of times.
\qpar

The case of Japan would
suggest that, in a general way, 
CV(birth) decreases strongly in the course 
of time. As a matter of fact such a conclusion would appear
fairly natural for one may think that in former times
sexual relations (and conceptions) were shaped by climatic
conditions, cultural traditions and religious precepts 
much more strongly
than they are nowadays. However, to our surprise, no matter
how natural, this idea was not found consistent with
observation. Switzerland offers a clear counter-example.
In the time interval, 1878--1885 CV(birth) is as low as 3.6\%.   
In subsequent years it increases to 9.0\% in 1920--1923
and falls back to 4.1\% in 2010-2013.
\qpar

In conclusion to this discussion one can retain that the
pattern of birth numbers is heavily influenced by exogenous
factors which, in addition, appear to be country-dependent.
\qpar 
 
In what follows we will try to answer a 
more limited question, namely
is the birth response to a
death peak of given magnitude the same no matter
whether the later is unexpected or on the contrary
a regular occurrence.
%
\begin{figure}[htb]
\centerline{\psfig{width=17cm,figure=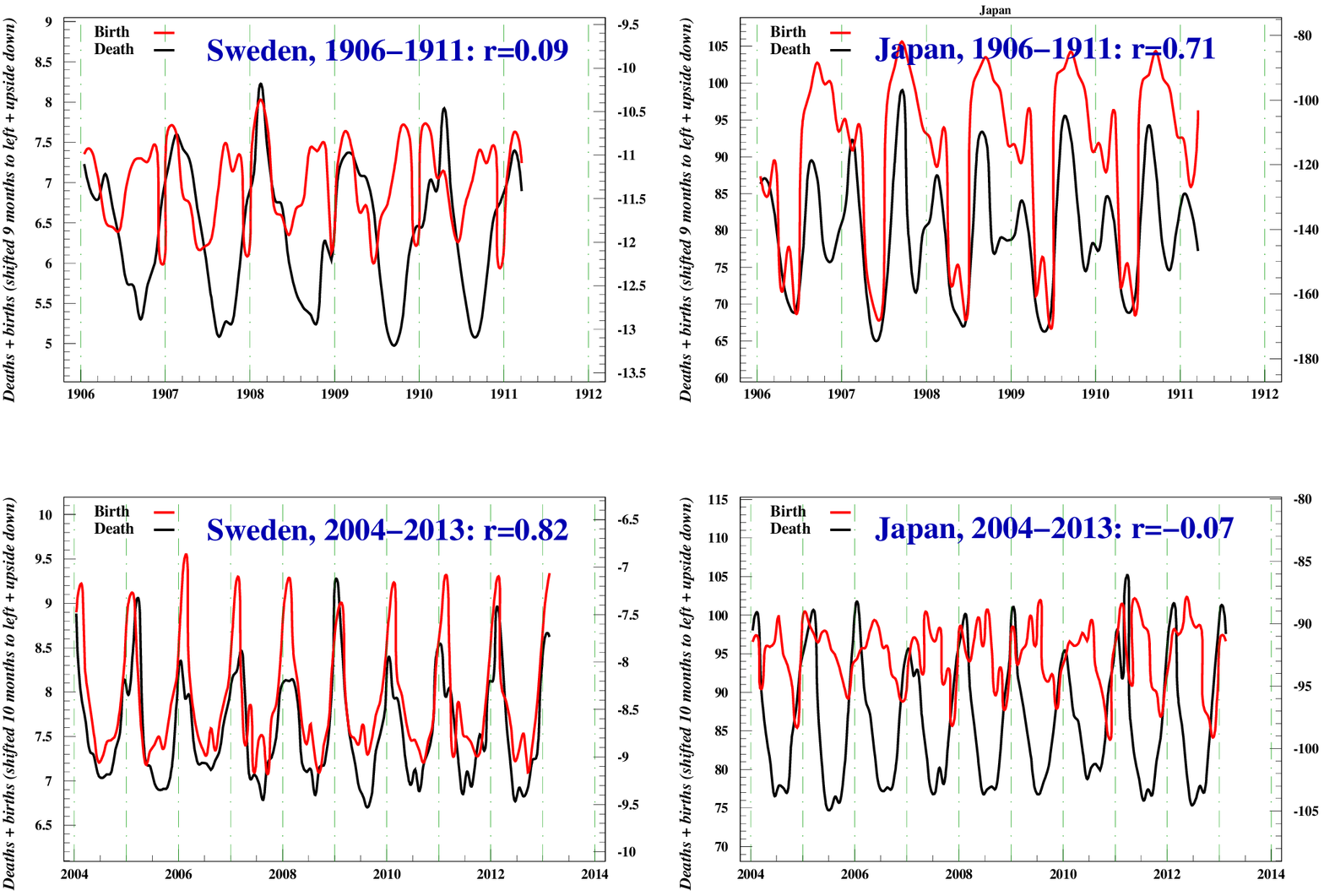}}
\qleg{Fig.\qhu 4a,b,c,d\qhv Is there a correlation between
seasonal death peaks and conception troughs in the early 20th
and 21st centuries?\qL}
{{\bf 1906-1911}: The death curves are very different in the two countries:
in Sweden there is a winter peak whereas in Japan there
is a summer peak. The conception curves are also very
different. For instance,
in Sweden there is a sharp conception spike
at the end of each year which may be due to 
Christmas time.\qL
{\bf 2004-2013}: Whereas the death peaks in Sweden and Japan are 
very similar, the conception curves are very different.
In Sweden the peaks of the inverted conception
curve coincide closely with the death spikes
which results in a high correlation between the two series.
On the contrary in Japan, the death and birth series
are disconnected which results in a correlation close
to zero. This shows that there is no systematic 
connection between deaths and births. The high synchronization
observed in Sweden cannot be considered as the rule and
is certainly due to special circumstances. This is
confirmed by the fact that other cases (e.g. Switzerland) 
are intermediate between the extreme cases of Sweden and Japan.\qL}
{Sources: 1906-1911: Bunle (1954); 2004-2013: 
Website of the Statistical Division of the United
Nations.}
\end{figure}

\qpar

%
\begin{table}[htb]

\small

\centerline{\bf Table 2 \quad Characteristics of ``normal'' 
seasonal fluctuations of
death and birth numbers}

\vskip 5mm
\hrule
\vskip 0.7mm
\hrule
\vskip 2mm

$$ \matrix{
\hbox{Country} \hfill & \hbox{Period} & \hbox{}\hfill &
\hbox{Coeff. of} & \hbox{Correlation}\cr
\qtb
\hbox{} \hfill & \hbox{} & \hbox{}\hfill &
\hbox{variation} & \hbox{death--birth(tr-inv)}\cr
\noalign{\hrule}
\qth
\hbox{Japan} \hfill & 1906-1911 & \hbox{}\hfill & & \cr
\hbox{} \hfill &  & \hbox{Death}\hfill & 10\% &  0.71\cr
\hbox{} \hfill &  & \hbox{Birth}\hfill & 22\% &  \cr
\hbox{Sweden} \hfill & 1906-1911 & \hbox{}\hfill & & \cr
\hbox{} \hfill &  & \hbox{Death}\hfill & 12\% &  0.09\cr
\hbox{} \hfill &  & \hbox{Birth}\hfill & 4.4\% &  \cr
\hbox{Switzerland} \hfill & 1906-1911 & \hbox{}\hfill & & \cr
\hbox{} \hfill &  & \hbox{Death}\hfill & 15\% &  0.77\cr
\hbox{} \hfill &  & \hbox{Birth}\hfill & 5.1\% &  \cr
\hbox{} \hfill &  & \hbox{}\hfill & & \cr
\hbox{Japan} \hfill & 2004-2013 & \hbox{}\hfill & & \cr
\hbox{} \hfill &  & \hbox{Death}\hfill & 9.3\% & -0.39 \cr
\hbox{} \hfill &  & \hbox{Birth}\hfill & 2.7\% &  \cr
\hbox{Sweden} \hfill & 2004-2013 & \hbox{}\hfill & & \cr
\hbox{} \hfill &  & \hbox{Death}\hfill & 7.8\% &  0.73\cr
\hbox{} \hfill &  & \hbox{Birth}\hfill & 7.1\% &  \cr
\hbox{Switzerland} \hfill & 2004-2013 & \hbox{}\hfill & & \cr
\hbox{} \hfill &  & \hbox{Death}\hfill & 8.6\% &  0.06\cr
\qtb
\hbox{} \hfill &  & \hbox{Birth}\hfill & 3.6\% &  \cr
\noalign{\hrule}
} $$
\vskip 1.5mm
Notes: The term ``normal'' in the title of the table means 
that no exceptional death spike occurred in the time intervals
under consideration.
The coefficient of variation is defined as the standard
deviation divided by the average. In the definition of the 
correlation, ``birth(tr-inv)'' means that the birth data have
been translated 9 months toward the past and inverted (i.e.
replaced by their opposite) in conformity with the graphs
drawn in the paper. Here the error bars on CV are less than 20\%
of the results.
{Sources: Website of the United Nations, Statistical Division}
\vskip 2mm
\hrule
\vskip 0.7mm
\hrule
\end{table}
%
What analytical tool should be used to answer this question?
The intercorrelation may be the first idea which comes to mind
but it is not satisfactory for an obvious reason.
We wish to single out the responses to death rate peaks
whereas the linear correlation will also reflect the response 
to death troughs or to flat death rates.
In addition, because of the attenuation, no visible
birth coupling should be expected when the death
peak is too small.
Despite its limitation the correlation can give valuable
information in two opposite cases: 
\qbu A correlation of 0.70 or higher
cannot be obtained if the peaks do not coincide. 
\qbu A negative correlation will indicate
that the peaks do not coincide.
\qpar

Naturally, this argument holds only under two conditions.
\qee{i} The amplitude of the seasonal death fluctuations must
no be too small
for otherwise, even if the effect exists,
it will too small to be detected at birth level.
In order to test this argument we use the methodology
of extreme cases that is to say, we compare two cases,
one in which the CV of the birth series is small
and another in which it is large (say about 30\%).
In addition we require that the two cases occur approximately
in the same time window in order to ensure similar 
environment conditions. 
\qpar

What conclusions can one draw from the results given in Table 2
and Fig.4?
\qee{i} The CV of the death series are fairly stable 
at a level of about 8\% both in time and across countries.
\qee{ii} The CV of the birth series are fairly different from
country to country; thus, in 1906-1911 they range from
4.4\% in Sweden to 22\% in Japan.  Contrary to our expectation
based on the argument given in the text above, the CV do not, as a rule,
decrease in the course of time. It is true that in Japan
there is a considerable decrease but in Sweden the CV increases
from 4.4\% to 7.1\%. 
\qee{iii} It is in the correlation that we are 
most interested. The cases with zero or negative
correlation do not necessarily imply
that the death spikes do not trigger birth troughs. This is
shown by Sweden (1906--1911). In this case the winter death
spike trigger birth troughs but in addition there are
major birth troughs in fall which are not triggered
by a death spike.
However, in Japan (2004--2013) although the death spikes are of 
larger amplitude than in Sweden  they do not trigger 
birth troughs. 
The conception time series is not at all in sync with the
winter peaks of the deaths.
\qpar

How can one explain that, despite this disconnection,
in some cases the correlation 
is fairly high? Our explanation is that this occurs purely
by chance. One can give a fairly crude argument.
For present-time data one can safely assume that the death series
has only one spike which occurs in winter time usually in January
or February%
\qfoot{In warm countries and former times there may also be
a death peak in summer time due in particular to enteritis
of babies and children.}%
.
On average this peak has a width of about 3 months.
If, as is the case for Sweden (2004--2013), the birth series
has also only one peak (and therefore one trough) 
then they may more or less
overlap with probability $ 3/12=0.25 $.
On the contrary, if the birth series has a more complex structure,
for instance with two peaks (and therefore two troughs), then
the single death peak can be in sink only with one of the 
birth troughs
which will result in a low correlation, as seen for Japan in 
2004--2013

\qA{Possible origin of the Japan-Sweden discrepancy}

As observed at the beginning of Paper 1,
 ``explanations''
relying on randomness are often a way to hide our lack
of understanding. So, let us assume for a moment that
the synchronicity observed in Sweden (2003--2014) is 
not purely due to chance. Where will such an assumption
lead us?
\qpar
As emphasized above, the graph of Japan (2004--2013) clearly
shows that there are cases where seasonal death 
spikes of an amplitude
exceeding 10\% of the mean fail to trigger conception troughs.
How can this be explained?
\qpar
It seems reasonable to assume that
in 2004-2013 both in Sweden and in Japan
the winter death peaks (i.e. zones 1+2) comprise mostly
elderly persons.
In accordance with what was said in Paper 1 and at the beginning
of the present paper, we then examine the broader 
set of persons in the 20-35 age interval who are affected
to some degree, i.e. zone $ H_{3,4}=H_3+H_4 $ (in the
notations of Paper 1, Fig. 2b).
 
$ H_{34} $ will comprise persons mildly
affected by winter diseases plus persons who grieve a lost family
member. If we assume that because of similar
health care systems the $ H_3 $ sets are basically the same
in the two countries
we are left with the conclusion that the grievance set
is much larger in Sweden than in Japan. On account of what
we know about close inter-generational links in Japan (as 
well as in China or Korea) such a conclusion appears surprising.
May be $ H_4 $ comprises other categories of 
persons that we did not consider so far? This is left
as an open question.

\qI{Conclusion}

It is the low level of noise which permitted the 
detection and analysis
of the coupling effect between death spikes and birth troughs.
Actually this statement must be qualified by saying
that, if irregular, the
seasonal fluctuations can be a major source of
noise but fortunately in many countries (and in particular
in Japan) they are sufficiently regular to be treated
as being deterministic signals%
\qfoot{In contrast
suicide rates have also a substantial seasonal component
but it is much less regular than the birth and death signals
considered here.}%
.
\qpar

As always when fairly accurate measurements are possible,
they raise a number of questions. How can one explain
that there is no coupling for 9/11 or for the winter
death spikes in Japan? Semi-quantitative
explanations based on the zone model (illustrated by Fig. 2b
of Paper 1) were proposed but they need to be confirmed
by additional evidence. If one could get monthly birth
and death data at province level for large countries
like China, India or Indonesia that would certainly allow
further progress.

\vskip 10mm
{\bf References}

\qparr
Bertillon (J.) 1892: La grippe \`a Paris et dans quelques
autres villes de France et de l'\'etranger en 1889--1890
[The influenza epidemic in Paris and in some other cities
in western Europe].
Imprimerie Municipale, Paris.\qL
[Available on Internet, for instance at the following address:\qL
{\small 
http://www.biusante.parisdescartes.fr/histoire/medica/resultats/index.php?do=livre\&
cote=20955}

\qparr
Bunle (H.) 1954: Le mouvement naturel de la population dans le
monde de 1906 \`a 1936. [Vital statistics of
many countries world-wide from 1906 to 1936.]
Editions de l'Institut d'Etudes D\'emographiques, Paris.

\qparr
Bureau of the Census [various years, starting in 1915]: Birth Statistics.
Government Printing Office, Washington DC.

\qparr
Bureau of the Census [various years]: Mortality Statistics.
Government Printing Office, Washington DC.

\qparr
Finland 1902. The French subtitle of this official
publication is: \'El\'ements d\'emo\-gra\-phi\-ques principaux de la
Finlande pour les ann\'ees 1750-1890, II: Mouvement de la population.
The Finnish title is:  Suomen [Finland] V\"aest\"otilastosta
[demographical elements] vuosilta [years] 1750-1890, II:
V\"aest\"on [population] muutokset [changes]. Helsinki 1902. \qL
[Available on line, 526 p.]

\qparr
Flora (P.), Kraus (F.), Pfenning (W.) 1987: State, economy, and
society in western Europe 1815-1975. A data handbook in two
volumes. Macmillan Press, London.

\qparr
Richmond (P.), Roehner (B.M.) 2017: Coupling between death
spikes and birth troughs. Part 1: Evidence.
Preprint LPTHE, UPMC, Paris.

\qparr
Roehner (B.M.) 2007: Driving forces in physical,
biological and socio-economic phenomena. A network
investigation of social bonds and interactions.
Cambridge University Press, Cambridge.

\qparr
Statistique G\'en\'erale de la France 1907: Statistique internationale
du mouvement de la population d'apr\`es les registres 
d'\'Etat Civil. R\'esum\'e r\'etrospectif depuis l'origine des
statistiques de l'\'Etat Civil jusqu'en 1905. 
[International vital statistics according to civil registry statistics.
Retrospective from the earliest times until 1905.]
Imprimerie Nationale, Paris.\qL
[This volume was published in the series 
``Statistique G\'en\'erale de la France'' but in fact it covers all
countries for which data were available in the early 20th century.]

\qparr
Sweden 1863. The Swedish title of this official publication is:
Befolknings-Statisk [population statistics] II. 1.
(Underd\aa niga Ber\"attelse) f\"or \aa ren 1856 med 1860.
Statistika Central-Byr\aa ns [Statistical Central Agency].
Stockholm, 1863. 

\end{document}